\documentclass[noshowpacs,pre,aps,superscriptaddress,floatfix,twocolumn]{revtex4}

\usepackage{lineno,hyperref,mathtools,doi}
\usepackage{amsmath}
\usepackage{amssymb,mathrsfs}
\usepackage{graphicx}
\usepackage{hyperref}

\DeclareMathAlphabet{\mathpzc}{OT1}{pzc}{m}{it}
\newcommand{\prt}{\partial}
\newcommand{\sss}{\scriptscriptstyle}

\begin{document}

\title{Long-time evolution of pulses in the Korteweg-de Vries equation in the
  absence of solitons revisited: Whitham method}

\author{M. Isoard}
 \affiliation{LPTMS, CNRS, Univ. Paris-Sud,
Universit\'e Paris-Saclay, 91405 Orsay, France}
\author{A. M. Kamchatnov} \affiliation{Institute of Spectroscopy,
  Russian Academy of Sciences, Troitsk, Moscow, 108840, Russia}
\affiliation{Moscow Institute of Physics and Technology, Institutsky
  lane 9, Dolgoprudny, Moscow region, 141701, Russia}
\author{N. Pavloff}
\affiliation{LPTMS, CNRS, Univ. Paris-Sud,
Universit\'e Paris-Saclay, 91405 Orsay, France}

\begin{abstract}
  We consider the long-time evolution of pulses in the Korteweg-de
  Vries equation theory for initial distributions which produce no soliton,
  but instead lead to the formation of a dispersive shock wave and of
  a rarefaction wave. An approach based on Whitham modulation theory
  makes it possible to obtain an analytic description of the structure
  and to describe its self-similar behavior near the soliton edge of
  the shock. The results are compared with numerical simulations.
\end{abstract}

\maketitle

\section{Introduction}

It is well known that pulses propagating through a nonlinear medium typically
experience wave breaking. Their long time evolution depends on which
effect---in addition to the nonlinearity---dominates after the wave
breaking moment: viscosity or dispersion. If viscosity dominates, then
the shock corresponds to a region of localized extend in which the
slow variables display a sharp transition. A typical small-amplitude
viscous shock can be modeled by the Burgers equation
\begin{equation}\label{eq1}
  u_t+uu_x=\nu u_{xx},
\end{equation}
for which a full analytic theory has been developed (see, e.g.,
Ref. \cite{whitham-74}).  For a positive initial profile
$u(x,t=0)\equiv u_0(x)>0$ which is well enough localized (i.e.,
$u_0(x)\to 0$ fast enough for $|x|\to\infty$) the time-evolved pulse
acquires a triangle-like shape at its front edge (or at its rear edge
if $u_0(x)<0$\footnote{A so-called $N$-wave appears if $u_0(x)$ has
  both polarities.}) gradually spreading out with decreasing
amplitude.

The situation changes drastically if dispersive effects dominate
rather than viscosity. In this case the typical evolution can be
described by the celebrated Korteweg-de Vries (KdV) equation
\begin{equation}\label{eq2}
  u_t+6uu_x+ u_{xxx}=0,
\end{equation}
which admits oscillating solutions ranging from linear waves to bright
solitons.  A positive localized initial pulse $u_0(x)>0$, after an
intermediate stage of wave breaking and complicated deformation,
eventually evolves into a sequence of solitons with some amount of
linear dispersive waves. The characteristics of the solitons are determined
by the initial distribution $u_0(x)$. If this initial pulse is intense
enough---so that the number of solitons is large---one may use for
determining the parameters of these solitons the asymptotic formula of
Karpman \cite{karpman-67} obtained in the framework of the inverse
scattering transform method discovered by Gardner, Green, Kruskal and
Miura~\cite{GGKM-67}.  However, if $u_0(x)<0$, since Eq.~\eqref{eq2}
does not admit dark (i.e., ``negative'') soliton, wave breaking does
not result in the formation of solitons in the asymptotic regime
$t\to\infty$, but it rather leads to the formation of a dispersive shock
wave (DSW) connected to a triangle-like rarefaction wave which is the
remnant of the initial trough. The shape and the time evolution of
this oscillatory structure is highly nontrivial and considerable
efforts have been invested into its study.

In an early investigation of
Berezin and Karpman~\cite{bk-64} it was shown that the KdV equation
admits solutions of the form
\begin{equation}\label{eq4}
  u(x,t)=\frac1{t^{2/3}}\, f\left(\frac{x}{t^{1/3}}\right),
\end{equation}
and numerical simulations of these authors demonstrated that some
region of the evolving wave structure is indeed described by solutions of
type \eqref{eq4}.  The existence of such a region was confirmed by the
inverse scattering transform
method in Refs.~\cite{shabat-73,an-73} and its ``quasi-linear''
part was studied in Ref.~\cite{zm-76}. An extensive study of the
asymptotic evolution of the pulse in the absence of solitons was
performed in Ref.~\cite{as-77} where different characteristic parts of
the wave structure were distinguished and their main parameters were
calculated. However, in this reference, Ablowitz and Segur---who first
explicitly point to the formation of a dispersive shock wave---confined 
themselves to the analytic study of typical limiting cases
and explicit formulae for the whole dispersive shock wave region were
found much later \cite{dvz-94} with the use of a quite
involved analysis of the associated Riemann-Hilbert problem.  This
approach was developed further in Refs.~\cite{ekv-01,cg-09,cg-10} and
other papers.

Although the above mentioned approaches are mathematically strict, the
methods used are difficult and the theory developed has not found
applications to concrete problems related with other integrable
evolution equations. Since the question of evolution of pulses in the
absence of solitons is related with experiments in physics of water
waves \cite{hs-78,tkco-16}, Bose-Einstein condensates
\cite{kgk-04,hoefer-06} and nonlinear optics \cite{wkf-07,xu-17}, the
development of a simpler and more transparent physically approach is
desirable.  Such an approach, based on the Whitham theory of
modulations of nonlinear waves \cite{whitham-65}, was suggested long
ago by Gurevich and Pitaevskii \cite{gp-73} and since that time it has
developed into a powerful method with numerous applications (see,
e.g., the review article \cite{eh-16}).  Despite the facts that some
elements of the Whitham theory were used in Refs.~\cite{an-73,as-77}
and that the general solution of the solitonless initial value problem
has been obtained in Ref.~\cite{ek-93}, no asymptotic analysis has
been performed within Whitham's formalism, so that its relationship
with the previous results remained unclear.

The main goal of the present paper is to fill this gap and to apply
the Whitham theory to the description of the asymptotic evolution of
initial pulses in the small dispersion limit (or for wide pulses)
under the condition of absence of solitons. We show that the
combination of two ideas---self-similarity of the solution and
quasi-simple character \cite{gkm-89} of the dispersive shock 
wave---permits a asymptotic analysis of the solution. The relatively simple
theory developed in the present work should be useful in the analysis
of experiments devoted to the evolution of pulses of this type.

The paper is organized as follows. In Sec.~\ref{WGH} we present the
main aspects of Whitham theory and of the generalized hodograph method
applying to quasi-simple waves (following
Refs.~\cite{gkm-89,Gur91,Gur92,Kry92,wright,tian,ek-93}). In
Sec.~\ref{sol-edge} the application of the ideas of Ref.~\cite{gkm-89}
to the soliton edge of the DSW makes it possible to find the law of
motion of this edge and suggests its self-similar asymptotic behavior
consistent with Eq.~(\ref{eq4}).  In Sec.~\ref{selfsim} we perform the
large time asymptotic analysis of the rear (soliton) part of the
dispersive shock wave by the Whitham method within the
self-similarity assumption. This yields a surprisingly simple
derivation of the solution of Ref.~\cite{dvz-94}.  The description of
the DSW in its full range by the method of El and Khodorovskii
\cite{ek-93} is presented in a self-contained manner in
Sec.~\ref{full}. In this section we consider the time evolution of
several initial profiles illustrating the possible different behaviors
in the shock region and compare the theoretical results with numerical
simulations. We present our conclusions in Sec.~\ref{conclusion}.

\section{Whitham theory and the generalized hodograph method}\label{WGH}

\subsection{The smooth part of the profile}

We consider an initial pulse with non-positive profile
$u(x,t=0)=u_0(x)$ defined on finite interval of $x$ and having a
single minimum min$_{x\in\mathbb{R}} \{u_0(x)\}=-1$ (this value can be
changed by an appropriate re-scaling on $u$, $x$ and $t$). The initial
profile is assumed to be smooth (i.e., for finite pulses with the
length $x_0$ we assume $x_0\gg 1$), so that in a first stage of
evolution one can neglect dispersive effects. This amounts to replace
the KdV dynamics by the Hopf equation
\begin{equation}\label{ek3}
  r_t+6\,r\, r_x=0\; .
\end{equation}
We change notation here to mark the difference between $r(x,t)$, solution
of \eqref{ek3}, and $u(x,t)$ which is the global solution of the KdV
equation \eqref{eq2}.

The solution of the Hopf equation is well known and it is given in
implicit form in terms of functions inverse to $u_0(x)$. In the case
we consider $u_0(x)$ has a single minimum and the inverse
function is two-valued. We denote its two branches as $w^{\sss\rm
  A}(r)$ and $w^{\sss\rm B}(r)$, where the first function refers to
the part of the pulse to the left of its minimum and the second one to
its right. Then the solution of the Hopf equation is given by
the formulae
\begin{subequations}\label{ek4}
\begin{align}
  x-6\,r\,t=w^{\sss\rm A}(r) , \label{ek4a} \\
x-6\,r\,t=w^{\sss\rm B}(r) .\label{ek4b}
\end{align}
\end{subequations}
For example, in case of a parabolic initial pulse
\begin{equation}\label{ek1}
u_0(x)=\begin{cases}
4\,x(x+x_0)/x_0^2 & \mbox{for}\;\; -x_0\le x \le 0\; , \\
0 & \mbox{elsewhere}\; ,
\end{cases}
\end{equation}
the inverse functions are equal to
\begin{equation}\label{ek2}
\begin{cases}
w^{\sss\rm A}(r)=\frac{\displaystyle x_0}{\displaystyle 2}
\left(-1-\sqrt{1+r}\,\right), \\[2mm]
w^{\sss\rm B}(r)=\frac{\displaystyle x_0}{\displaystyle 2}
\left(-1+\sqrt{1+r}\,\right),
\end{cases}
\mbox{where}\;
r\in[-1,0] .
\end{equation}
\begin{figure}
\includegraphics*[width=\linewidth]{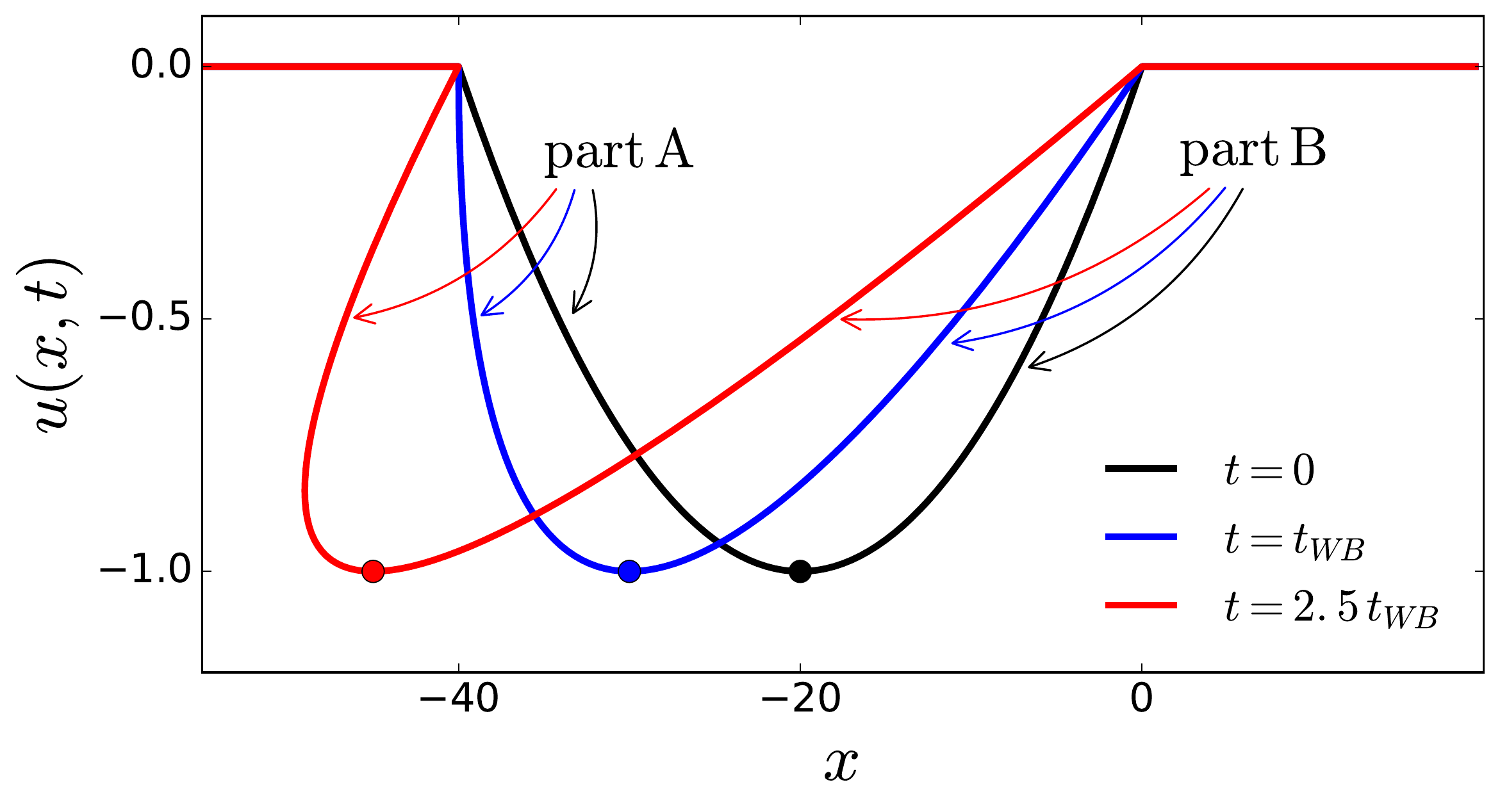}
\caption{Dispersionless evolution of the initial parabolic profile
  \eqref{ek1} with $x_0=40$. The black, blue and red solid lines
  represent $r(x,t)$ solution of \eqref{ek3} for times $t=0$,
  $t=t_{\sss\rm WB}$ and $t=2.5\, t_{\sss\rm WB}$. The dots represent
  the position of the minimum min$_{x\in\mathbb{R}} \{r(x,t)\}$ which
  separates parts A (at the left) and B (at the right) of the
  profile.}\label{fig1}
\end{figure}
\begin{figure}
\includegraphics*[width=\linewidth]{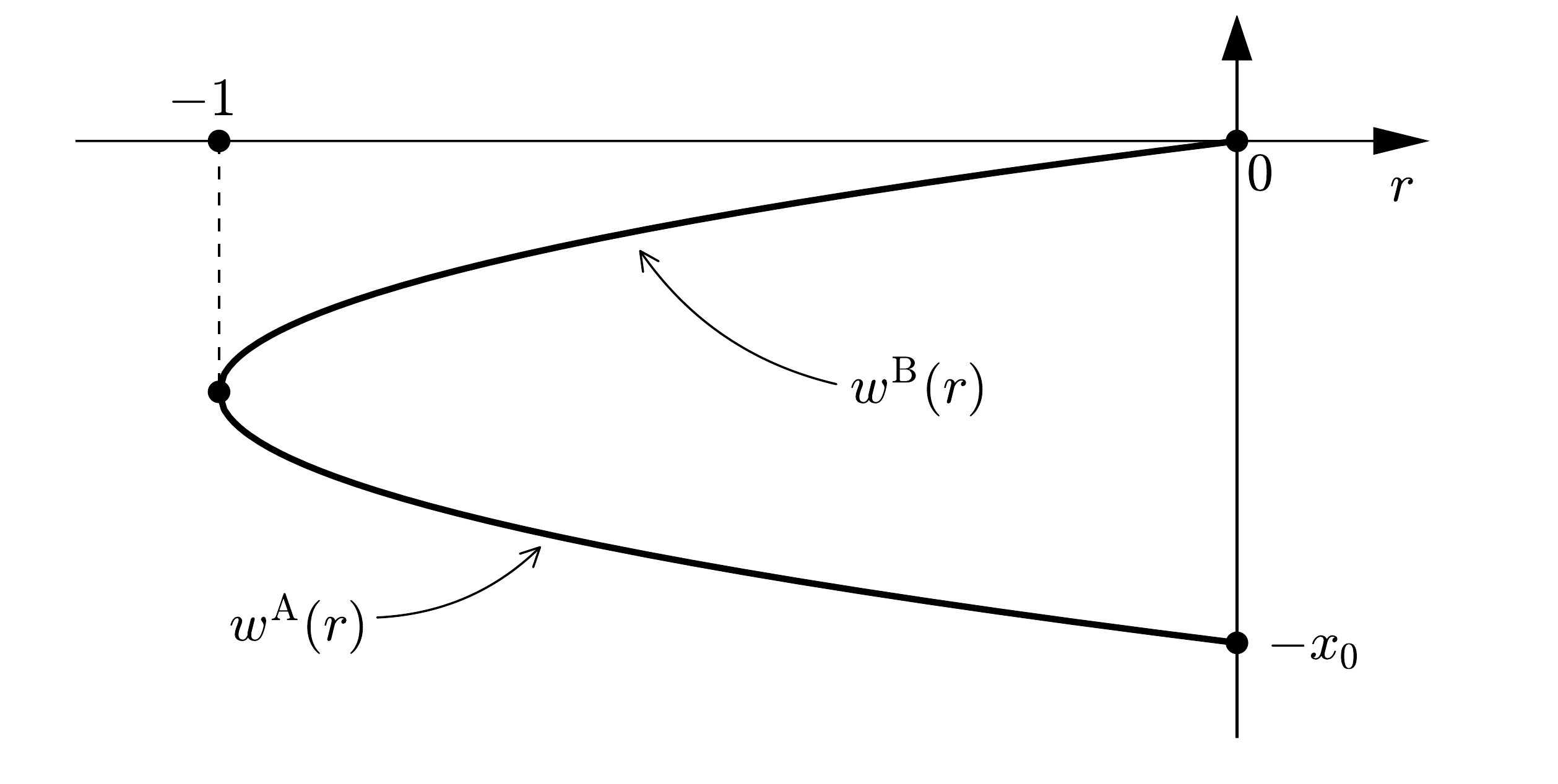}
\caption{The two branches $w^{\sss\rm A}(r)$ and $w^{\sss\rm B}(r)$ of
  the reciprocal function of $u_0(x)$. The figure is drawn for the
  initial parabolic profile \eqref{ek1} (the corresponding expressions
  of $w^{\sss\rm A}(r)$ and $w^{\sss\rm B}(r)$ are given in
  Eq.~\eqref{ek2}) but the behavior is the generic one.}\label{figwAB}
\end{figure}
Figure \eqref{fig1} represents the initial profile \eqref{ek1} and its
time evolution as computed from Eqs.~\eqref{ek4}, i.e., without taking
dispersive effects into account. Figure \eqref{figwAB} represents the
corresponding functions $w^{\sss\rm A}(r)$ and $w^{\sss\rm B}(r)$.  In
the following we shall illustrate the explicit computations by this
simple initial profile. Other types of profiles, with less generic
behaviors, will be presented and discussed in Sec. \ref{full}.

The wave breaking time is the time
$t_{\sss\rm WB}=1/{\rm max}(-6 \, {\rm d}u_0/{\rm d}x)$ at which the
solution of \eqref{ek3} becomes infinitely steep (see, e.g.,
Ref.~\cite{whitham-74}). In the present work we consider initial
profiles for which the largest slope
${\rm max} (-{\rm d}u_0/{\rm d}x)$ is reached at $x=-x_0$ for $r=0$ and thus
\begin{equation}\label{eq:wb}
t_{\sss\rm WB}=-\frac{1}{6} \left(\frac{d w^{\sss\rm A}}{dr}\right)_{r=0}\; .
\end{equation}
 For the initial profile \eqref{ek1} we get
$t_{\sss\rm WB}=x_0/24$. For $t\ge t_{\sss\rm WB}$ the dispersionless
approximation fails (the corresponding formal solution of the Hopf
equation is multi-valued), and a DSW is formed, initially around
$x=-x_0$, which then propagates in the negative $x$ direction. We now
explain how it can be described within Whitham modulational theory.

\subsection{Periodic solutions and their modulations}\label{sec:DSW}

The KdV equation \eqref{eq2} admits nonlinear periodic solutions which
can be written in terms of three parameters $r_1\leq r_2\leq r_3$ as
(see, e.g., \cite{kamch-2000})
\begin{equation}\label{ek9}
\begin{split}
  u(x,t)=& r_3+r_2-r_1-2(r_2-r_1)\times\\
& \mathrm{sn}^2(\sqrt{r_3-r_1}(x-Vt),m),\\
\mbox{where}\;\;  V=&2(r_1+r_2+r_3),\quad\mbox{and}\quad
m=\frac{r_2-r_1}{r_3-r_1}.
  \end{split}
\end{equation}
The notation ``sn'' in the above formula refers to the Jacobi sine
function (see, e.g., Ref.~\cite{gr}). For constant parameters $r_i$
expression \eqref{ek9} is an exact (single phase) solution of the KdV
equation, periodic in time and space with wavelength
\begin{equation}\label{ek16}
  L=\frac{2K(m)}{\sqrt{r_3-r_1}},
\end{equation}
where $K(m)$ is the complete elliptic integral of the first kind.

According to the Gurevich-Pitaevskii scheme, a DSW may be described as
a modulated nonlinear periodic wave of type \eqref{ek9} for which the
$r_i$'s slowly depend on time and position and evolve according to the
Whitham equations (see, e.g, Refs.~\cite{kamch-2000,eh-16})
\begin{equation}\label{ek14}
  \prt_t r_i+v_i(r_1,r_2,r_3)\prt_x r_i=0,
\qquad i=1,2,3.
\end{equation}
The quantities $v_i$ in these equations are the Whitham velocities. Their
explicit expressions have first been derived by Whitham
\cite{whitham-65}, and can also be obtained from the relation
\begin{equation}\label{ek15}
  v_i=\left(1-\frac{L}{\prt_{i} L}\, \prt_i\right) L=
V -2\, \frac{L}{\prt_{i}L}\; ,
\end{equation}
where $L$ is the wavelength \eqref{ek16} of the nonlinear periodic
solution \eqref{ek9} and $\prt_i$ stands for $\prt_{r_i}$. One
gets
\begin{equation}\label{eq12ak}
  \begin{split}
  &v_1=2(r_1+r_2+r_3)+\frac{4(r_2-r_1)K(m)}{E(m)-K(m)},\\
  &v_2=2(r_1+r_2+r_3)-\frac{4(r_2-r_1)(1-m)K(m)}{E(m)-(1-m)K(m)},\\
  &v_3=2(r_1+r_2+r_3)+\frac{4(r_3-r_1)(1-m)K(m)}{E(m)},
  \end{split}
\end{equation}
where $E(m)$ is the complete elliptic integral of the second kind.

Since Eqs.~(\ref{ek14}) have a diagonal form (that is, they include
derivatives of a single parameter $r_i$ in each equation), the
variables $r_i$ are called Riemann invariants of the Whitham 
equations---Riemann was the first who introduced such variables 
in the theory of nonlinear waves.

The two edges of the DSW are denoted as $x_{\sss\rm L}(t)$ and
$x_{\sss\rm R}(t)$. The first one is the small amplitude edge, it is
at the left of the DSW in the case we consider.  Within Whitham
approximation, it makes contact between the DSW and the undisturbed profile:
$u(x,t)=0$ for $x\le x_{\sss\rm L}(t)$. The small amplitude version of
\eqref{ek9} corresponds to the limit $m\ll 1$ and takes the form
\begin{equation}\label{ek10}
  u(x,t)=r_3+(r_2-r_1)\cos[2\sqrt{r_3-r_1}(x-V t)].
\end{equation}
In this harmonic linear limit, $r_2\to r_1$ ($m\to0$) and the Whitham
velocities (\ref{eq12ak}) reduce to
\begin{equation}\label{eq14ak}
\begin{split}
&  \left.v_1\right|_{r_2=r_1}=\left.v_2\right|_{r_2=r_1}=12\,r_1-6\,r_3,\\
&   \left.v_3\right|_{r_2=r_1}=6\,r_3.
\end{split}
\end{equation}
Around the left boundary of the DSW,
the amplitude $2(r_2-r_1)$ of the oscillations is small and
since this edge propagates along a zero background, we arrive at the
conclusion that $r_3=0$ and $r_1=r_2$ for $x=x_{\sss\rm L}(t)$.

The other edge, at the right side of the DSW, is the large amplitude
soliton edge, with $m=1$. Therefore we must have here $r_2=r_3$ and
in this limit the nonlinear pattern \eqref{ek9}
degenerates into a soliton solution of the form
\begin{equation}\label{ek11}
  u(x,t)=r_1+
\frac{2(r_2-r_1)}{\cosh^2[\sqrt{r_2-r_1}(x-V t)]}\; .
\end{equation}
This implies that the right of the DSW is bounded by a soliton for which the
Whitham velocities are given by
\begin{equation}\label{eq13ak}
\begin{split}
& \left.v_1\right|_{r_2=r_3}=6\, r_1,\\
&  \left.v_2\right|_{r_2=r_3}=\left.v_3\right|_{r_2=r_3}=2\, r_1+4\, r_3.
\end{split}
\end{equation}
The contact of the DSW with the smooth profile (which prevails for
$x\ge x_{\sss\rm R}(t)$) imposes the condition
$r_1(x_{\sss\rm R}(t),t) = r(x_{\sss\rm R}(t),t)$,
where $r(x,t)$ is a solution of the Hopf equation \eqref{ek3} with
initially $r(x,0)=u_0(x)$.

Therefore the description of the DSW for
$x\in[x_{\sss\rm L}(t),x_{\sss\rm R}(t)]$ is consistent with a
constant value $r_3(x,t)=0$ for the larger Riemann parameter,
while the two others satisfy the boundary conditions
\begin{subequations}\label{ek12}
\begin{align}
\begin{split}
&  r_1(x_{\sss\rm R}(t),t) = r(x_{\sss\rm R}(t),t) \equiv r_{\sss\rm R}(t),\\
& r_2(x_{\sss\rm R}(t),t) = 0, \label{ek12a}
\end{split}
\\
&r_1(x_{\sss\rm L}(t),t) = r_2(x_{\sss\rm L}(t),t)\equiv r_{\sss\rm L}(t) .
\label{ek12b}
\end{align}
\end{subequations}
Note that all the above functions are only defined after the wave
breaking time, i.e., for $t\ge t_{\sss\rm WB}$.

\subsection{Generalized hodograph method}\label{sec:GHM}

As just discussed, for the type of structure we aim at describing, two
Riemann invariants ($r_1$ and $r_2$) change along the DSW. The corresponding
shock is thus not a simple wave solution corresponding to a step-like
initial profile; it belongs to the
class of ``quasi-simple waves'' introduced in Ref.~\cite{gkm-89}.  In
this case, Eq.~\eqref{ek14} with $i=3$ is trivially satisfied (by
$r_3=0$) and for solving the remaining two Whitham equations we use
the so-called generalized hodograph method of Tsarev \cite{Tsa91}.  To
this end, one introduces two functions $W_i(r_1,r_2)$ ($i=1$ or 2)
making it possible to write a vector generalization of Eq.~\eqref{ek4}
for the Whitham system
\begin{equation}\label{ek17}
x-v_i(r_1,r_2) t =W_i(r_1,r_2), \quad i=1,2\; .
\end{equation}
For the sake of brevity we have noted in the above equation
$v_i(r_1,r_2)=v_i(r_1,r_2,r_3=0)$ for $i\in\{1,2\}$; we will keep this
notation henceforth. The $W_i$'s must satisfy the compatibility
equation found by substituting \eqref{ek17} into \eqref{ek14}. This
leads to the Tsarev equations:
\begin{equation}\label{ek18}
\frac{\prt_{j} W_i}{W_i-W_j}=\frac{\prt_{j} v_i}{v_i-v_j}\; ,
\; \mbox{for}\; i\neq j\; .
\end{equation}
One can show (see, e.g., \cite{Gur92,wright,tian}) that
\eqref{ek18} is solved for $W_i$'s of the form
\begin{equation}\label{ek19}
W_i = \left(1-\frac{L}{\prt_{i}L}\prt_{i}\right)\mathscr{W}
=\mathscr{W}+(\tfrac12 v_i -r_1-r_2) \prt_{i} \mathscr{W}\; ,
\end{equation}
where $\mathscr{W}(r_1,r_2)$ is solution of the Euler-Poisson equation
\begin{equation}\label{ek20}
\partial_{12}\mathscr{W}=
\frac{\prt_1\mathscr{W}-\prt_2\mathscr{W}}{2(r_1-r_2)}\; .
\end{equation}
There is however a subtle point here, which was first understood in
Ref.~\cite{gkm-89} (see also Ref.~\cite{ek-93}): after the wave
breaking time, the development of the dispersive shock wave occurs in
{\it two} steps. Initially (when $t$ is close to $t_{\sss\rm WB}$),
the DSW is connected at its right edge to the smooth profile coming
from the time evolution of part A of the initial profile. We denote
this as case ``A'' which occurs in ``region A'' of the $(x,t)$ plane.
Then, after a while, the left part of the initial profile (part A) has
been ``swallowed'' by the DSW which is then connected to the smooth
profile coming from the time evolution of part B of $u_0(x)$ (this is
case ``B'', ``region B'' of the $(x,t)$ plane). In case A, for a given
time $t$, the lower value of $u(x,t)$ is reached within the smooth
part of the profile and keeps its initial values ($-1$). In case B,
the minimum ${\rm min}_{x\in \mathbb{R}} \{u(x,t)\}$ is reached inside
the DSW (or at its boundary), is negative and larger than $-1$ (i.e., less
pronounced than in case A) and asymptotically tends to 0 for large
time.

In region A of the $(x,t)$ plane, we denote by
$\mathscr{W}^{\sss\rm A}(r_1,r_2)$ the solution of the Euler-Poisson
equation, in region B we denote it instead as
$\mathscr{W}^{\sss\rm B}(r_1,r_2)$. These two forms are joined by the line
$r_1=-1$ (cf. the upper left plot of Fig.~\ref{fig3}) where
\begin{equation}\label{ek21}
\mathscr{W}^{\sss\rm A}(-1,r_2)=\mathscr{W}^{\sss\rm B}(-1,r_2)\; .
\end{equation}
Since the general solution of the Euler-Poisson equation with the
appropriate boundary conditions, and the construction of the resulting
nonlinear pattern are quite involved, we shall first consider some
particular---but useful---results which follow from general principles of
the Whitham theory.

\section{Motion of the soliton edge of the shock}\label{sol-edge}

During the first stage of evolution of the DSW, its right (solitonic)
edge is connected to the smooth dispersionless solution described by
formula (\ref{ek4a}), that is we have here
\begin{equation}\label{eq1ak}
  x_{\sss\rm R}-6r_{\sss\rm R}t=w^{\sss\rm A}(r_{\sss\rm R}).
\end{equation}
On the other hand, in vicinity of this boundary the Whitham equations
(\ref{ek14}) with the limiting expressions (\ref{eq13ak}) (where
$r_3=0$) for the velocities $v_i$ are given by
\begin{equation}\label{t3-148.1}
  {\prt_t}r_1+6r_1{\prt_x}r_1=0,\qquad
  {\prt_t}r_2+2r_1{\prt_x}r_2=0.
\end{equation}
For solving these equations one can perform a classical hodograph
transformation (see, e.g., \cite{kamch-2000}), that is, one assume
that $x$ and $t$ are functions of the independent variables $r_1$ and
$r_2$: $t=t(r_1,r_2)$, $x=x(r_1,r_2)$. We find from Eqs.~\eqref{t3-148.1} 
that these functions must satisfy the linear system
\begin{equation}\nonumber
    \frac{\prt x}{\prt r_1}-2r_1\frac{\prt t}{\prt r_1}=0,\qquad
    \frac{\prt x}{\prt r_2}-6r_1\frac{\prt t}{\prt r_2}=0.
\end{equation}
At the boundary with the dispersionless solution [where
$r_1=r_{\sss\rm R}$, see \eqref{ek12a}] the first equation reads
\begin{equation}\label{t3-148.3}
  \frac{\prt x_{\sss\rm R}}{\prt r_{\sss\rm R}}
-2\,r_{\sss\rm R}\frac{\prt t}{\prt r_{\sss\rm R}}=0,
\end{equation}
and this must be compatible with Eq.~(\ref{eq1ak}). Differentiation of
Eq.~(\ref{eq1ak}) with respect to $r_{\sss\rm R}$ and elimination of
$\prt x_{\sss\rm R}/\prt r_{\sss\rm R}$ with the use of
Eq.~\eqref{t3-148.3} yield the differential equation for the function
$t(r_{\sss\rm R})=t(r_{\sss\rm R},0)$:
\begin{equation}\label{t3-148.5}
  4\,r_{\sss\rm R}\frac{dt}{dr_{\sss\rm R}}+6\,t=
-\frac{dw^{\sss\rm A}(r_{\sss\rm R})}{d r_{\sss\rm R}}.
\end{equation}
At the wave breaking time $r_{\sss\rm R}=0$ and \eqref{t3-148.5} gives
the correct definition \eqref{eq:wb} of the wave breaking
time: $t_{\rm\sss WB}=t(0)$. Elementary integration then yields
\begin{equation}\label{t3-148.6}
\begin{split}
  t(r_{\sss\rm R})& =\frac1{4(-r_{\sss\rm R})^{3/2}}
\int_0^{r_{\sss\rm R}}\!\!\!\sqrt{-r}\,\frac{dw^{\sss\rm A}(r)}{d r}\, dr\\
& = \frac1{8(-r_{\sss\rm R})^{3/2}}
\int_0^{r_{\sss\rm R}} \frac{w^{\sss\rm A}(r)}{\sqrt{-r}}dr
-\frac{w^{\sss\rm A}(r_{\sss\rm R})}{4\,r_{\sss\rm R}}.
\end{split}
\end{equation}
Substituting this expression into (\ref{eq1ak}), we get the
function $x_{\sss\rm R}(r_{\sss\rm R})=x(r_{\sss\rm R},0)$,
\begin{equation}\label{t3-148.7}
\begin{split}
  x_{\sss\rm R}(r_{\sss\rm R})& =-\frac3{2\sqrt{-r_{\sss\rm R}}}
\int_0^{r_{\sss\rm R}}\!\!\!\sqrt{-r}\,
\frac{dw^{\sss\rm A}(r)}{d r}dr+w^{\sss\rm A}(r_{\sss\rm R})\\
& = -\frac3{4\sqrt{-r_{\sss\rm R}}}
\int_0^{r_{\sss\rm R}} \frac{w^{\sss\rm A}(r)}{\sqrt{-r}}dr
-\frac{1}{2}w^{\sss\rm A}(r_{\sss\rm R}).
\end{split}
\end{equation}
The two formulae \eqref{t3-148.6} and \eqref{t3-148.7} define in an
implicit way the law of motion $x=x_{\sss\rm R}(t)$ of the soliton
edge of the DSW.

The above expressions are correct as long as the soliton edge is located
inside region A, that is up to the moment
\begin{equation}\label{t3-148.6ak}
  t_{\sss\rm A/B}=t(-1)=\frac1{4}\int_0^{-1}\sqrt{-r}\,
\frac{dw^{\sss\rm A}(r)}{d r}dr,
\end{equation}
after which the soliton edge connects with region
B. Concretely, for a time $t>t_{\sss\rm A/B}$, we have to solve the
differential equation
\begin{equation}\nonumber
  4\,r_{\sss\rm R}\frac{dt}{dr_{\sss\rm R}}+6\, t=
-\frac{dw^{\sss\rm B}(r_{\sss\rm R})}{d r_{\sss\rm R}}
\end{equation}
with the initial condition $t(-1)=t_{\sss\rm A/B}$. This yields
\begin{equation}\label{t3-148.8a}
\begin{split}
  t(r_{\sss\rm R})=&\frac1{4(-r_{\sss\rm R})^{3/2}}
\Big(\int_0^{-1}\sqrt{-r}\,\frac{dw^{\sss\rm A}(r)}{d r}dr\\
  &+\int_{-1}^{r_{\sss\rm R}}\sqrt{-r}\,\frac{dw^{\sss\rm B}(r)}{d r}dr\Big),
\end{split}
\end{equation}
and
\begin{equation}\label{t3-148.8b}
\begin{split}
  x_{\sss\rm R}(r_{\sss\rm R})=&-\frac3{2(-r_{\sss\rm R})^{1/2}}
\Big(\int_0^{-1}\sqrt{-r}\,\frac{dw^{\sss\rm A}(r)}{d r}dr\\
&+\int_{-1}^{r_{\sss\rm R}}\sqrt{-r}\,\frac{dw^{\sss\rm B}(r)}{d r}dr\Big)
+w^{\sss\rm B}(r_{\sss\rm R}).
  \end{split}
\end{equation}

At asymptotically large time $t\to\infty$ one in at stage B of
evolution with furthermore $r_{\sss\rm R}\to0$.  Hence the upper limit
of integration in the second integrals of formulae \eqref{t3-148.8a}
and \eqref{t3-148.8b} can be put equal to zero. Integration over $r$
in the resulting expressions can be replaced by integration over $x$
with account of the fact that $w^{\rm\sss A,B}(r)$ represent two
branches of the inverse function of $r=u_0(x)$, so we get
\begin{equation}\nonumber
  t(r_{\sss\rm R})\approx\frac{\cal A}{4(-r_{\sss\rm R})^{3/2}},
\quad\mbox{where}\quad
  {\cal A}=\int_{\mathbb{R}}\sqrt{-u_0(x)}\, dx\
\end{equation}
is a measure of the amplitude of the initial trough.
Consequently,
\begin{equation}\label{xrrr}
 r_{\sss\rm R}(t)=-\left(\frac{\cal A}{4 t}\right)^{2/3}\; ,\quad
 x_{\sss\rm R}(t)=-\frac{3{\cal A}^{2/3}}{2^{1/3}}t^{1/3}\; ,
\end{equation}
where we have neglected $x_0$ which is small compared with the infinitely
increasing time-dependent terms.

At large time, the dispersionless part of the profile between $x=0$
and $x_{\sss\rm R}(t)$ is stretched to a quasi-linear behavior
$u(x,t)=x\,r_{\sss\rm R}(t)/x_{\sss\rm R}(t)$, and one thus has
\begin{equation}
\int_{x_{\sss\rm R}(t)}^0 dx \sqrt{-u(x,t)}=-\frac{2}{3}
x_{\sss\rm R}(t) \sqrt{-r_{\sss\rm R}(t)}={\cal A}\; ,
\end{equation}
which means that the quantity ${\cal A}$ is conserved, at least at the
level of the present asymptotic analysis. This situation is
reminiscent of---but different from---the dissipative case where
nonlinear patterns of triangular shape may also appear at the rear
edge of a (viscous) shock. In the dissipative case also there exists a
conserved quantity. For Burgers equation for instance, with an initial
condition of type \eqref{ek1}, a single viscous shock appears which is
followed by an asymptotically triangular wave. This means that the
details of the initial distribution are lost (as in the present case)
but for Burgers equation the conserved quantity is $\int_{I(t)}\!dx\, u(x,t)$,
where $I(t)$ is the support of the triangular wave (equivalent to our
segment $[x_{\sss\rm R}(t),0]$).

Formulae (\ref{xrrr}) suggest that in the vicinity of the soliton
edge, the behavior of the DSW must be self-similar, and we now turn to the
investigation of this possibility in the framework of Whitham theory.

\section{Similarity solution at the soliton edge of the
  shock}\label{selfsim}

In this section we use the Whitham approach to obtain the
long time asymptotic behavior of the shock close to
$x_{\sss\rm R}(t)$, valid up to $x\sim - t^{1/3} (\ln t)^{3/2}$
(see Refs. \cite{as-77,dvz-94}).

\begin{figure}
\includegraphics*[width=\linewidth]{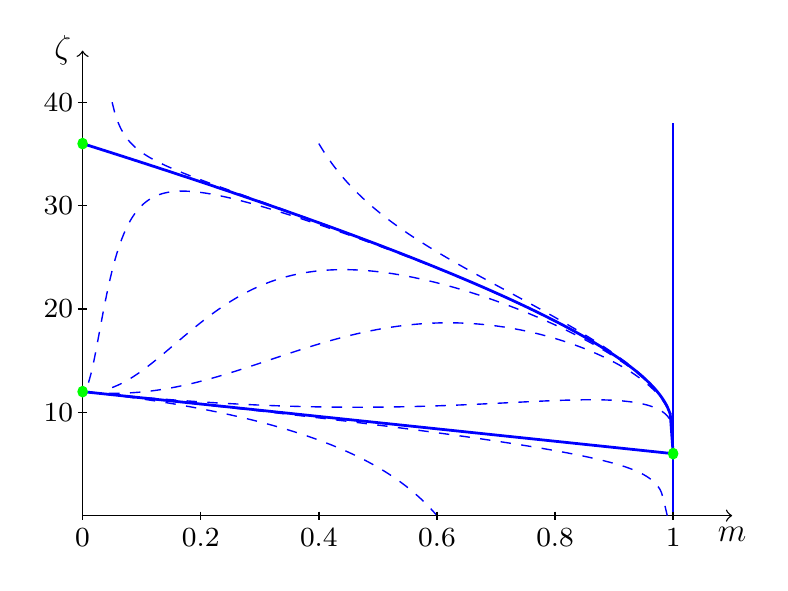}
\caption{Integral curves of Eq.~\eqref{eq22a}. The separatices are
depicted as solid thick lines. }
\label{fig2a}
\end{figure}

Equations \eqref{xrrr} suggest that, close to the soliton edge of the DSW,
the Riemann invariants $r_1$ and $r_2$ have the following scaling form:
\begin{equation}\label{39.2}
  r_i=\frac1{t^{2/3}}{R}_i
\left(\frac{x}{t^{1/3}}\right).
\end{equation}
Here $x<0$ and since $r_1<r_2<0$, we have $R_1<R_2<0$. The scaling
\eqref{39.2} agrees with the scaling (\ref{eq4}) of the full KdV
equation first noticed in Refs.~\cite{bk-64,shabat-73,an-73}. Written
in terms of the re-scaled Riemann parameters $R_1$ and $R_2$ and of
the self-similar variable $z=x/t^{1/3}$, the Whitham equations
\eqref{ek14} read
\begin{equation}\label{eq18}
  \frac{dR_i}{dz}=-\frac{2R_i}{z-3R_1V_i(m)},\qquad i=1,2,
\end{equation}
where
\begin{equation}\label{eq20}
  m=1-{R_2}/{R_1},
\end{equation}
and the velocities $V_1(m)$ and $V_2(m)$ are given by
\begin{equation}\label{eq19}
  \begin{split}
  &V_1(m)=2(2-m)-\frac{4mK(m)}{E(m)-K(m)},\\
  &V_2(m)=2(2-m)+\frac{4m(1-m)K(m)}{E(m)-(1-m)K(m)}.
  \end{split}
\end{equation}
The two equations (\ref{eq18}) can be reduced to a single equation
if we introduce the variable
\begin{equation}\label{39.6}
  \zeta={z}/{R_1}
\end{equation}
and look for the dependence of $\zeta$ on $m$. Simple calculation yields
the differential equation
\begin{equation}\label{eq22a}
  \frac{d\zeta}{dm}=
\frac{(\zeta-V_1(m))(\zeta-3V_2(m))}{2(1-m)(V_2(m)-V_1(m))}
\end{equation}
whose basic properties can be studied in the phase plane $(m,\zeta)$.
The phase portrait in this plane is displayed in
Fig.~\ref{fig2a}.
It has the singular points
\begin{equation}\nonumber
  \begin{array}{llll}
  & (0,12),\qquad &(0,36)\qquad &\text{for}\qquad m=0;\\
  & (1,6),\qquad &(1,6)\qquad &\text{for}\qquad m=1,
  \end{array}
\end{equation}
that is at $m=1$ two singular points merge into one of a mixed type:
for a part of the integral curves it is a saddle point and for the
other part it is a source. Numerical solution of Eq.~(\ref{eq22a})
suggests that the separatrix joining the singular points $(0,12)$ and
$(1,6)$ is a straight line
\begin{equation}\label{t3-149.6}
  \zeta=6(2-m),
\end{equation}
which, after returning to the variables $R_1,R_2$ and $z$ leads to the
assumption that the system (\ref{eq18}) admits the following integral:
\begin{equation}\label{t3-149.7}
  R_1+R_2=\frac16z
\end{equation}
A direct check shows that indeed
$d(R_1+R_2)/dz=1/6$ under the condition (\ref{t3-149.7}), so that this
assumption is proved. The integral curves beginning in vicinity of
this separatrix are attracted to it when $m$ decreases, so one can
expect that just this separatrix realizes the self-similar regime of
the DSW near its soliton edge.

To determine the dependence of $m$ on $z$, we find, with the use of
Eqs.~(\ref{eq18}),
\begin{equation}\label{t3-149.8}
  \frac{dm}{dz}=
\frac{6\zeta(m)(1-m)(V_2(m)-V_1(m))}{z[\zeta(m)-3V_1(m)][\zeta(m)-3V_2(m)]}.
\end{equation}
Substituting Eq.~(\ref{t3-149.6}) and the expressions (\ref{eq19}) in
the above, we get the following equation
\begin{equation}\label{t3-150.9}
  \frac{dm}{dz}=-\frac{2-m}{zmK(m)} F(m),
\end{equation}
where
\begin{equation}\label{t3-150.9b}
F(m)=(2-m)E(m)-2(1-m)K(m).
\end{equation}
The solution of this equation determines $m=m(z)$ along the separatrix.

The form of expression \eqref{t3-150.9b} suggests that it can be
obtained as a result of the calculation of some elliptic integral in which
the integration limits may play the role of more convenient
variables. Inspection of tables of such integrals shows that the
formula 3.155.9 in Ref.~\cite{gr} (which we write down here with
notations slightly different from the original reference),
\begin{equation}\label{Iint}
\begin{split}
I= & 3\int_{q_2}^{q_1}\sqrt{(q_1^2-y^2)(y^2-q_2^2)}\,dy\\
=  & q_1\left[(q_1^2+q_2^2)E(\mathpzc{m})-2q_2^2K(\mathpzc{m})\right],
\end{split}
\end{equation}
has a necessary structure.  In Eq.~\eqref{Iint} one has $q_1>q_2>0$
and $\mathpzc{m}=1-(q_2/q_1)^2$.

To establish the link between the
two expressions \eqref{t3-150.9b} and \eqref{Iint}, it is enough to
take
\begin{equation}\label{t3-150.10b}
q_1^2+q_2^2=1,
\end{equation}
so that $1-\mathpzc{m}=q_2^2/q_1^2$, $2-\mathpzc{m}=1/q_1^2$. Assuming that
the variables $q_1,q_2$ satisfy (\ref{t3-150.10b}), we obtain
\begin{equation}\label{t3-150.11}
  q_1^2=\frac1{2-\mathpzc{m}},\qquad q_2^2=\frac{1-\mathpzc{m}}{2-\mathpzc{m}},
\end{equation}
and then, imposing $\mathpzc{m}=m$ we get $F(m)=(2-m)^{3/2} I$.

Since $dq_1/dm=q_1^3/2$, Eq.~\eqref{t3-150.9} can be cast under the form
\begin{equation}\label{t3-150.13}
  \frac{dq_1}{d\ln(-z)}=-\frac{q_1}{2mK(m)} \, F(m),
\end{equation}
which is more convenient for further calculations. On the other hand,
the integral (\ref{Iint}) with account of Eqs.~(\ref{t3-150.11})
simplifies to
$$
I=q_1[E(m)-2(1-q_1^2)K(m)],
$$
and its differentiation with respect to $q_1$ gives
\begin{equation}\label{t3-150.15}
  \frac{dI}{dq_1}=3q_1^2mK(m).
\end{equation}
With the help of the formulae obtained we transform Eq.~(\ref{t3-150.13}) to
$$
\frac{dI}{I}=-\frac32d\ln(-z).
$$
Then, integration of this equation with the boundary condition
$z=z_1$ at $m=1$ yields $z$ as a function of $m$:
\begin{equation}\label{t3-150.17}
z=z_1\, I^{-2/3}(m)=z_1\, \frac{2-m}{F^{2/3}(m)}
\end{equation}
where
\begin{equation}\label{t3-150.17z1}
z_1=\frac{x_{\sss\rm R}(t)}{t^{1/3}}=-6 \left({\cal A}/{4}\right)^{2/3}
\end{equation}
is the value of $z$ for $m=1$ (at the soliton edge of the DSW, see
sec.~\ref{sol-edge}).

From the formulae $m=1-R_2/R_1$ and (\ref{t3-150.11}) we find the relationship
between $R_i$ and $q_i$:
\begin{equation}\label{t3-150.19}
  R_1=\frac{q_1^2}{6}\,z,\qquad R_2=\frac{q_2^2}{6}\,z,
\end{equation}
so that for the dependence of the Riemann invariants on $m$ we obtain
\begin{equation}\label{t3-150.20}
\begin{split}
  & R_1(m)=\frac{z_1}{6(2-m)I^{3/2}},\\
  & R_2(m)=\frac{(1-m)z_1}{6(2-m)I^{3/2}}.
  \end{split}
\end{equation}
Formulae \eqref{t3-150.17}, \eqref{t3-150.17z1} and \eqref{t3-150.20},
together with Eq.~(\ref{39.2}), completely determine the self-similar
solution of the Whitham equations: for fixed $t$ we have
$x(m)=t^{1/3}z(m)$, so that all functions are defined parametrically,
with $m$ playing the role of the parameter. Up to notations, this
solution coincides with the one obtained in Ref.~\cite{dvz-94} by
means of the study of asymptotic Riemann-Hilbert problem.

In the harmonic limit $m\ll1$, the relation \eqref{t3-150.17} reads
\begin{equation}\label{eq31}
  m=m_1z^{-3/4},\quad\mbox{where}\quad
m_1=\frac{2^{11/4}}{\sqrt{3\pi}}\,(-z_1)^{3/4},
\end{equation}
which leads to the expressions
\begin{equation}\label{eq32}
  r_1=\frac{x}{12t}-\frac{m_1}{24}\frac{(-x)^{1/4}}{t^{3/4}},\quad
  r_2=\frac{x}{12t}+\frac{m_1}{24}\frac{(-x)^{1/4}}{t^{3/4}}.
\end{equation}
It is important to notice that the difference $r_2-r_1$, that is, the
amplitude of the oscillations in the ``quasilinear'' region of Zakharov
and Manakov~\cite{zm-76}, increases with growing distance from the
soliton edge [as $(-x)^{1/4}$], but $r_2/r_1\to1$ and $m\to0$ here.
Hence, this limit is not a small-amplitude one and therefore the
self-similar regime cannot be realized along the whole DSW; it
takes place close enough to the soliton edge only. Consequently,
we have to turn to the general solution of the Whitham equations
to obtain the full description of the DSW.

\section{General solution}\label{full}

In this section, following Ref.~\cite{ek-93}, we turn to the general
solution of the Whitham equations given by the formulae of
Sec.~\ref{sec:GHM}. Our task now is to express the functions
$W_i(r_1,r_2)$, $i=1,2$, in terms of the initial form $u_0(x)$ of the
pulse. As was indicated above, at the first stage of evolution the DSW
is located inside the region A and after the moment $t_{\sss\rm A/B}$
[see Eq.~(\ref{t3-148.6ak})] a second stage begins where it reaches
region B. Correspondingly, the expressions for $W_i$ ans $\mathscr{W}$
are given by different formulae and should be considered separately.

\subsection{Solution in region A}

In region A one can follow the procedure explained in
Ref.~\cite{Gur92}. One imposes the matching of the right edge of the
DSW with the dispersionless solution \eqref{ek4}: just at
$x=x_{\sss\rm R}(t)$, we have $r_1=r(x,t)$, where $r(x,t)$ is the
solution of \eqref{ek3}, and $v_1(r_1,0)=6\, r_1$ (this follows from
Eq.~\eqref{eq13ak}). Comparing in this case Eqs.~\eqref{ek4} and
\eqref{ek17} one obtains
\begin{equation}\label{ek22}
W^{\sss\rm A}_1(r_1,0)=w^{\sss\rm A}(r_1)\; ,
\end{equation}
which embodies the same information as Eq.~\eqref{ek12a}. In terms of
$\mathscr{W}$ this corresponds to the equation
\begin{equation}\label{ek23}
\mathscr{W}^{\sss\rm A}(r_1,0) + 2\, r_1\,
\partial_1 \mathscr{W}^{\sss\rm A}(r_1,0) = w^{\sss\rm A}(r_1)\; ,
\end{equation}
whose solution is
\begin{equation}\label{ek24}
\mathscr{W}^{\sss\rm A}(r_1,0)= \frac{1}{2\sqrt{-r_1}}
\int_{r_1}^0 \frac{w^{\sss\rm A}(\rho)\, {\rm d}\rho}{\sqrt{-\rho}}\; .
\end{equation}
This will serve as a boundary condition for the
Euler-Poisson equation \eqref{ek20} whose general solution has
been given by Eisenhart \cite{Eis18} in the form
\begin{equation}\label{ek25}
\begin{split}
\mathscr{W}^{\sss\rm A}(r_1,r_2)=  &\int_{r_1}^0
\frac{\varphi^{\sss\rm A}(\mu)\, {\rm d}\mu}{\sqrt{(\mu-r_1)|r_2-\mu|}} +\\
& \int_{r_2}^0
\frac{\psi^{\sss\rm A}(\mu)\, {\rm d}\mu}{\sqrt{(\mu-r_1)(\mu-r_2)}}\; ,
\end{split}
\end{equation}
where the functions $\varphi^{\sss\rm A}$ and $\psi^{\sss\rm A}$ are
arbitrary functions to be determined from the appropriate boundary
conditions.  By taking $r_2=0$ in this expression one sees that
$\varphi^{\sss\rm A}(\mu)/\sqrt{-\mu}$ is the Abel transform of
$\mathscr{W}^{\sss\rm A}(r_1,0)$. The inverse transform reads
\cite{Ark05}
\begin{equation}\label{ek26}
\frac{\varphi^{\sss\rm A}(\mu)}{\sqrt{-\mu}}=
-\frac{1}{\pi} \frac{\rm d}{{\rm d}\mu}
\int_\mu^0 \frac{\mathscr{W}^{\sss\rm A}(r,0)\, {\rm d}r}{\sqrt{r-\mu}} \; .
\end{equation}
Plugging expression \eqref{ek24} for $\mathscr{W}^{\sss\rm A}(r,0)$ in
this formula and changing the order of integration one obtains
\begin{equation}\label{ek27}
\varphi^{\sss\rm A}(\mu)=\frac{1}{2\,\pi\sqrt{-\mu}}
\int_\mu^0 \frac{w^{\sss\rm A}(\rho)\,{\rm d}\rho}{\sqrt{\rho-\mu}} \; .
\end{equation}
For the initial profile \eqref{ek1}, $w^{\sss\rm A}$ is given in
Eq.~\eqref{ek2} and one gets explicitly
\begin{equation*}
\varphi^{\sss\rm A}(\mu)=-\frac{x_0}{4\pi}
\left(3+\frac{1+\mu}{\sqrt{-\mu}}\tanh^{-1} \sqrt{-\mu}\right).
\end{equation*}
In order to determine the function $\psi^{\sss\rm A}$, one considers
the left boundary of the DSW where, according to \eqref{ek12b}, $r_1$
and $r_2$ are asymptotically close to each other. Let us write $r_1=r$
and $r_2=r+\epsilon$ with $r\in[-1,0]$ and $\epsilon$ small and
positive. One gets from \eqref{ek25}
\begin{equation}\label{ek27b}
\begin{split}
\mathscr{W}^{\sss\rm A}(r,r+\epsilon) &=  \int_{r+\epsilon}^0\!\! {\rm d}\mu\,
\frac{\varphi^{\sss\rm A}(\mu)+\psi^{\sss\rm A}(\mu)}
{\sqrt{(\mu-r)(\mu-r-\epsilon)}}\\
& +\int_r^{r+\epsilon}
\frac{\varphi^{\sss\rm A}(\mu)\,{\rm d}\mu}{\sqrt{(\mu-r)(r+\epsilon-\mu)}}\; .
\end{split}
\end{equation}
In the right hand side of the above equality, the second term
converges when $\epsilon$ tends to 0 [towards $\pi\varphi^{\sss\rm
  A}(r)$], whereas the first one diverges unless $\varphi^{\sss\rm
  A}(r) +\psi^{\sss\rm A}(r)=0$, this being true for all
$r\in[-1,0]$. This imposes that the functions $\varphi^{\sss\rm A}$ and
$\psi^{\sss\rm A}$ should be opposite one the other and the final form of
the Eisenhart solution in case A reads
\begin{equation}\label{ek28}
\mathscr{W}^{\sss\rm A}(r_1,r_2)= \int_{r_1}^{r_2}
\frac{\varphi^{\sss\rm A}(\mu)\, {\rm d}\mu}{\sqrt{(\mu-r_1)(r_2-\mu)}}\; ,
\end{equation}
where $\varphi^{\sss\rm A}$ is given by formula \eqref{ek27}.

\subsection{Solution in region B}

One looks for a solution of the Euler-Poisson equation in
region B of the form
\begin{equation}\label{ek29}
\mathscr{W}^{\sss\rm B}(r_1,r_2)=\mathscr{W}^{\sss\rm A}(r_1,r_2)+
\int_{-1}^{r_1} \!
\frac{\varphi^{\sss\rm B}(\mu)\, {\rm d}\mu}{\sqrt{(r_1-\mu)(r_2-\mu)}}.
\end{equation}
Indeed, this ensures that $\mathscr{W}^{\sss\rm B}$, (i) being the sum of
two solutions of the Euler-Poisson equation, is also a solution of
this equation and (ii) verifies the boundary condition \eqref{ek21}
since the second term of the right-hand side of \eqref{ek29} vanishes
when $r_1=-1$.

At the right boundary of the DSW, $\mathscr{W}^{\sss\rm B}(r_1,0)$
verifies the same equation \eqref{ek23} as $\mathscr{W}^{\sss\rm A}(r_1,0)$
does, where all the superscripts A should be replaced by B. The
solution with the appropriate integration constant reads
\begin{equation}\label{ek30}
\begin{split}
\mathscr{W}^{\sss\rm B}(r_1,0) & = \frac{1}{2\sqrt{-r_1}}
\int_{r_1}^{-1} \frac{w^{\sss\rm B}(\rho)\, {\rm d}\rho}{\sqrt{-\rho}} \\
&+ \frac{1}{2\sqrt{-r_1}}
\int_{-1}^{0} \frac{w^{\sss\rm A}(\rho)\, {\rm d}\rho}{\sqrt{-\rho}}\; .
\end{split}
\end{equation}
The same procedure than the one previously used in part A of the DSW
leads here to
\begin{equation}\label{ek31}
\varphi^{\sss\rm B}(\mu)=\frac{1}{2\,\pi\sqrt{-\mu}}
\int_{-1}^\mu \!{\rm d}\rho\,
\frac{w^{\sss\rm A}(\rho)-w^{\sss\rm B}(\rho)}{\sqrt{\mu-\rho}} \; .
\end{equation}
For the initial profile \eqref{ek1} one gets explicitly
\begin{equation*}
\varphi^{\sss\rm B}(\mu)=-\frac{x_0}{4}\, \frac{1+\mu}{\sqrt{-\mu}}\; .
\end{equation*}
In the generic case, Eqs.~\eqref{ek29} and \eqref{ek31} give the
solution of the Euler-Poisson equation in region B.

\subsection{Characteristics of the DSW at its
  edges}\label{sec:Boundaries}

It is important to determine the boundaries $x_{\sss\rm R}(t)$ and
$x_{\sss\rm L}(t)$ of the DSW, as well as the values of the
Riemann invariants $r_1$ and $r_2$ at these points. The law of motion
of the soliton edge was already found in Sec.~\ref{sol-edge} and
it is instructive to show how this result can be obtained from the
general solution.

At the soliton edge we have $r_2=r_3=0$ and $r_1=r_{\sss\rm
  R}(t)$.
The corresponding Whitham velocities are $v_1=6\, r_{\sss\rm R}$ and
$v_2=2\, r_{\sss\rm R}$ [see Eqs. \eqref{eq13ak}], and the two
equations \eqref{ek17} read
\begin{equation}\label{ek32}
\begin{split}
x_{\sss\rm R}-6 r_{\sss\rm R} t =W_1(r_{\sss\rm R},0)& = w(r_{\sss\rm R}), \\
x_{\sss\rm R}-2 r_{\sss\rm R} t =W_2(r_{\sss\rm R},0)& =\mathscr{W}(r_{\sss\rm R},0).
\end{split}
\end{equation}
These formulae apply to both stages of evolution and therefore the
superscripts A and B are dropped out. They give at once
\begin{equation}\label{35.28}
  \begin{split}
  & t(r_{\sss\rm R})=\frac1{4r_{\sss\rm R}}
\left[\mathscr{W}(r_{\sss\rm R},0)-w(r_{\sss\rm R})\right],\\
  & x_{\sss\rm R}(r_{\sss\rm R})=
\frac12\left[3\mathscr{W}(r_{\sss\rm R},0)-w(r_{\sss\rm R})\right].
  \end{split}
\end{equation}
Let us consider the stage A for instance. Eq.~\eqref{ek24} yields
$$
\mathscr{W}^{\sss\rm A}(r_{\sss\rm R},0)=
- \frac1{2\sqrt{r_{\sss\rm R}}}\int^{r_{\sss\rm R}}_0
\frac{w^{\sss\rm A}(\rho)d\rho}{\sqrt{-\rho}},
$$
which, inserted into
Eqs.~(\ref{35.28}) gives immediately the results (\ref{t3-148.6}) and
(\ref{t3-148.7}).  For instance, for the initial profile \eqref{ek1},
when the right boundary is still in region A, one obtains explicitly
\begin{equation}\label{ek33}
t(r_{\sss\rm R})=\frac{x_0}{16\,r_{\sss\rm R}}
\left(\sqrt{1+r_{\sss\rm R}}-
\frac{\arcsin\sqrt{-r_{\sss\rm R}}}{\sqrt{-r_{\sss\rm R}}}\right)\; .
\end{equation}
At the wave breaking time $r_{\sss\rm R}=0$ and this yields
$t_{\sss\rm WB}=t(r_{\sss\rm R}=0)=x_0/24$ as already obtained
[cf. Eq.~\eqref{eq:wb}]. Stage A ends at time $t_{\rm\sss A/B}$ at which
the minimum (-1) of the smooth part of the profile enters the
DSW. This corresponds to $t_{\rm\sss A/B}=t(r_{\sss\rm R}=-1)$ and yields, for the
initial parabolic profile \eqref{ek1}: $t_{\rm\sss A/B}=\pi\, x_0/32$.

Let us now turn to the determination of the location $x_{\sss\rm L}(t)$
of the left boundary of the DSW, and of the common value
$r_{\sss\rm L}(t)$ of $r_1$ and $r_2$ at this point. In the typical
situation the left boundary is located in region A. In this case the
equations \eqref{ek17} for $i=1$ and 2 are equivalent and read
\begin{equation}\label{ek34}
x_{\sss\rm L}-12\, r_{\sss\rm L} \cdot t =W_1^{\sss\rm A}(r_{\sss\rm L},r_{\sss\rm L})\; .
\end{equation}
An equation for $r_{\sss\rm L}$ alone is obtained by demanding that
the velocity ${\rm d}x_{\sss\rm L}/{\rm d}t$ of the left boundary is
equal to the common value $12\, r_{\sss\rm L}$ of $v_1$ and $v_2$ at
this point [cf. Eqs.~\eqref{eq14ak}]. The time derivative of
Eq.~\eqref{ek34} then yields
\begin{equation}\label{ek35}
t =-\frac{1}{12}\,
\frac{{\rm d}W_1^{\sss\rm A}(r_{\sss\rm L},r_{\sss\rm L})}{{\rm d}r_{\sss\rm L}}\; .
\end{equation}
Once $r_{\sss\rm L}(t)$ has been determined by solving this equation,
$x_{\sss\rm L}(t)$ is given by Eq.~\eqref{ek34}.

Note that the relation ${\rm d}x_{\sss\rm L}/{\rm d}t=12\, r{\sss\rm L}$
is a consequence of the general statement that the small amplitude
edge of the DSW propagates with
the group velocity corresponding to the wave number determined by the
solution of the Whitham equations. Indeed, the KdV group velocity of
a linear wave with
wave-vector $k$ moving over a zero background is $v_g=-3 k^2$, and here
$k=2\pi/L=2\sqrt{-r_{\sss\rm L}}$ [cf. Eq.~\eqref{ek16}], hence
$v_g=12 \, r_{\sss\rm L}={\rm d}x_{\sss\rm L}/{\rm d}t$, as it should be.
This property of the small-amplitude edge is especially important
in the theory of DSWs for non-integrable equations (see \cite{El05,Kam18}).

We also study below a case different from \eqref{ek1} for which the
left boundary of the DSW belongs to region B and corresponds to
$r_1=r_2=-1$ [in the so-called triangular case corresponding to
$u_0(x)$ given by Eq.~\eqref{ek37}]. Then, at the small amplitude edge
$v_1=v_2=-12$ and Eqs.~\eqref{ek17} yield $x_{\sss\rm L}+12\cdot
t=C^{\rm st}$, the constant being the common value of $W_1^{\sss\rm
  B}(-1,-1)$ and $W_2^{\sss\rm B}(-1,-1)$. It can be determined at
$t=t_{\sss\rm WB}$, leading in this case to
\begin{equation}\label{ek36}
x_{\sss\rm  L}=-x_0-12 (t-t_{\sss\rm WB})\; .
\end{equation}
It is worth noticing that the velocity $dx_{\sss\rm  L}/dt=-12$ agrees with
the leading term in Eq.~(\ref{eq32}) for $r_1=-1$ in spite of a
non-vanishing amplitude of the self-similar solution in this limit.
For a more detailed study of the small-amplitude
region beyond the Whitham approximation see, e.g., Ref. \cite{cg-10}.

\subsection{The global picture}\label{sec:global}

\begin{figure}
\includegraphics*[width=\linewidth]{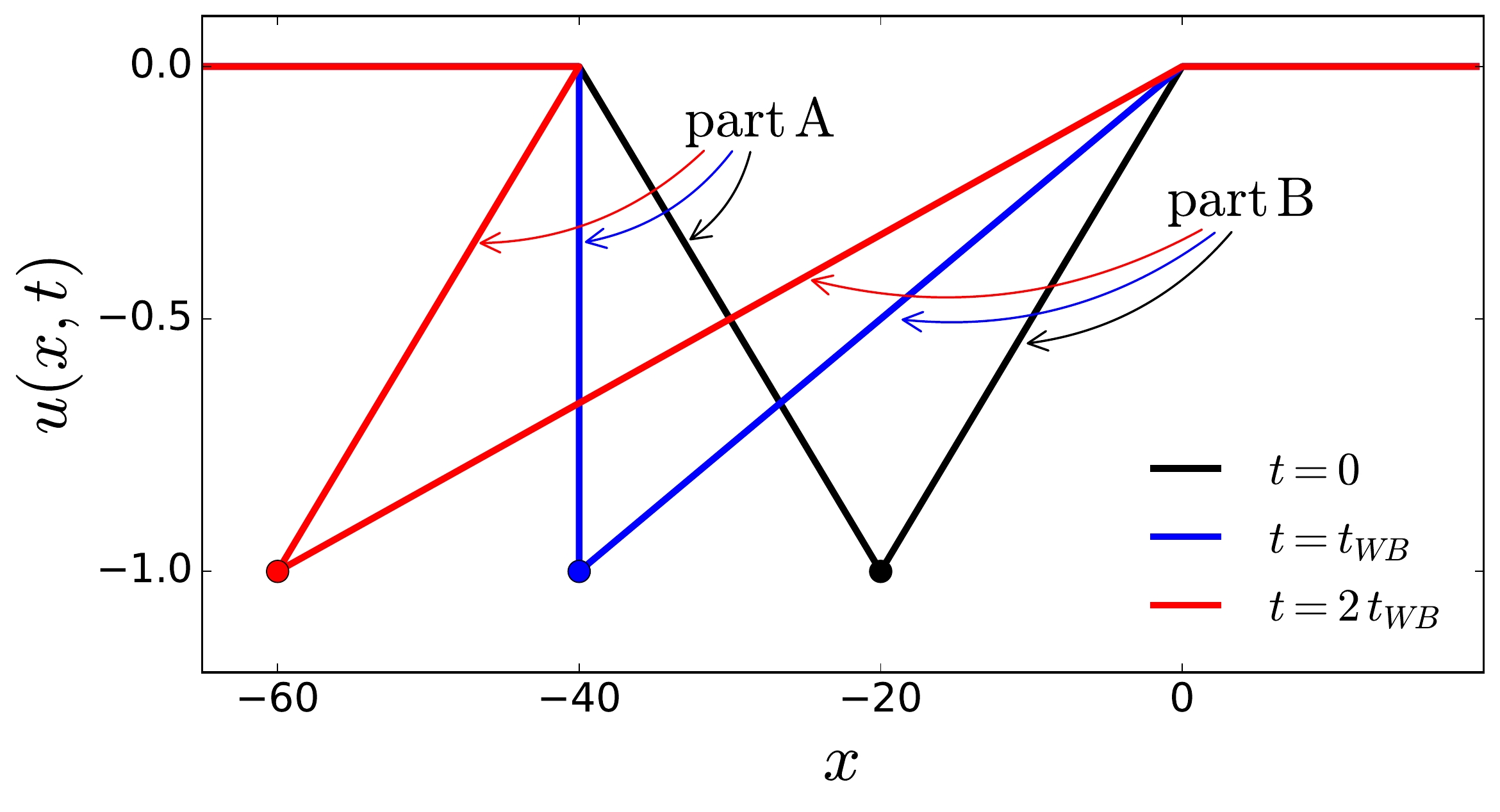}
\caption{Dispersionless evolution of the initial triangular profile
  \eqref{ek37} with $x_0=40$. The black, blue and red solid lines
represent $r(x,t)$ solution of \eqref{ek3}
  for times $t=0$, $t=t_{\sss\rm WB}$ and $t=2\, t_{\sss\rm
    WB}$.}\label{fig2}
\end{figure}

We now compare the results of the Whitham approach with the numerical
solution of the KdV equation for the initial profile \eqref{ek1} and
also for a profile
\begin{equation}\label{ek37}
u_0(x)=\begin{cases}
-1+\left|\frac{\displaystyle 2 x}{\displaystyle x_0}+1\right| &
\mbox{for}\quad -x_0\le x \le 0\; , \\
0 & \mbox{elsewhere.}
\end{cases}
\end{equation}
This profile is represented in Fig.~\ref{fig2} at $t=0$, at
wave-breaking time $t=t_{\sss\rm WB}$, which in the present case is
equal to $t_{\rm\sss WB}=x_0/12$, and also at $t=2\, t_{\sss\rm WB}$
(in the dispersionless approximation).

We henceforth denote the initial profile \eqref{ek1} as ``parabolic''
and the initial profile \eqref{ek37} as ``triangular''.
As was indicated above, the triangular
profile has the particularity of having a DSW within the region B only.
This is clear from Fig.~\ref{fig2}: part A of the initial profile does
not penetrate into the DSW region before part B does. Or, phrasing this
differently: according to the dispersionless evolution, at
$t=t_{\sss\rm WB}$ both parts A and B penetrate into the region of
multi-valuedness at $x\le x_0$.

\begin{figure*}[hptb]
\includegraphics*[width=1.9\columnwidth]{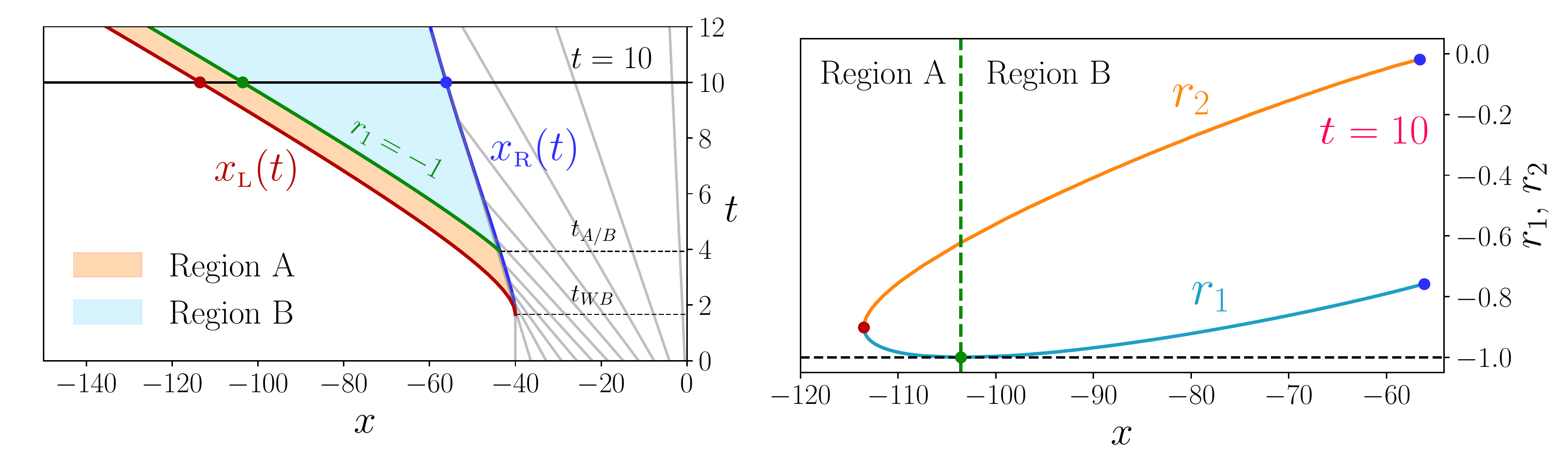}
\includegraphics*[width=1.9\columnwidth]{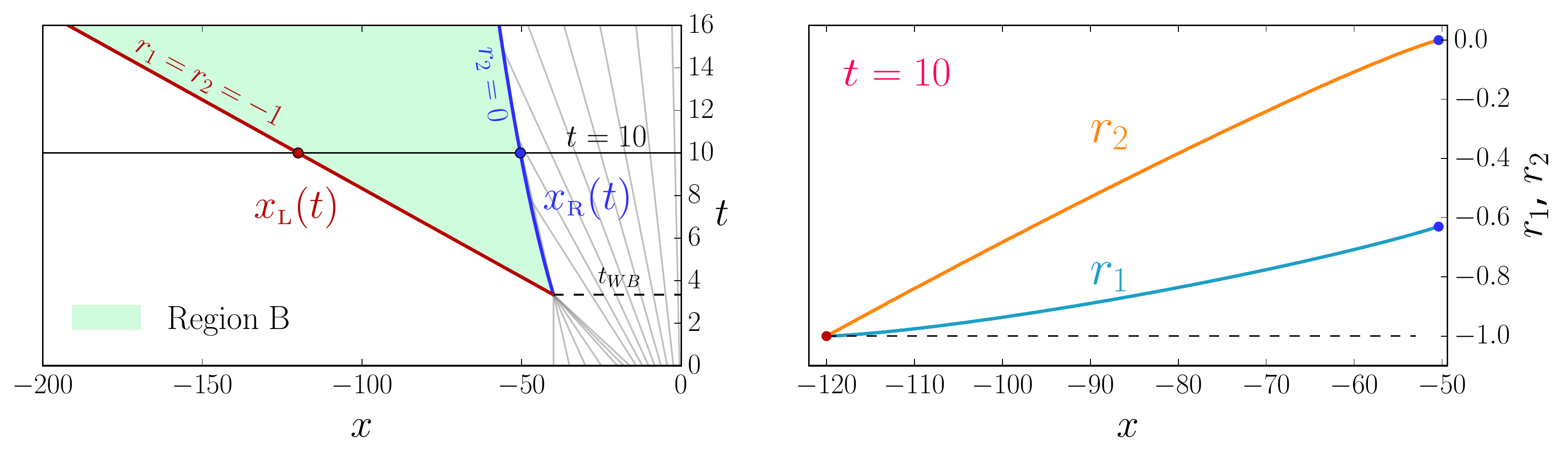}
\caption{The lower plots refer to the triangular initial profile, and
  the upper ones to the parabolic initial profile. Left column:
  different regions in the $(x,t)$ plane. The DSW occurs in the
  colored regions. The characteristics of the dispersionless evolution
  are represented as gray lines. In the upper left plot the time
  $t_{\sss\rm A/B}$ is the time where part A of the initial profile
  has been completely absorbed by the DSW. For the triangular profile
  $t_{\sss\rm A/B}=t_{\sss\rm WB}$. Right column: plot of the two
  varying Riemann invariants $r_1$ and $r_2$ at fixed $t=10$
  for $x_{\sss\rm L}(t)\le x \le x_{\sss\rm R}(t)$. }
  \label{fig3}
\end{figure*}

The DSW is described by Whitham method as explained in Sections
\ref{sec:DSW} and \ref{sec:GHM}. For this purpose one needs to
determine $r_1$ and $r_2$ as functions of $x$ and $t$ ($r_3\equiv
0$). This is performed as follows:
\begin{itemize}
\item First, we pick up a given $r_1 \in [-1, \,r_{\sss\rm R}]$, where
  $r_{\sss\rm R}$ is the value of $r_1$ at the soliton edge, the point
  where the DSW is connected to the rarefaction wave (it has been explained
  in Sec.~\ref{sec:Boundaries} how to compute it).

\item Second, at fixed $t$ and $r_1$, we find the corresponding value
  $r_2$ as a solution of the difference equation obtained from
  Eqs.~\eqref{ek17}
\begin{equation}\label{solve_r12}
\left( v_1 - v_2 \right)\cdot t = W_2(r_1,r_2) -W_1(r_1,r_2)
\, ,
\end{equation}
where $W_1$ and $W_2$ are computed from Eq.~\eqref{ek19}.
\item Last, the corresponding value of $x$ is determined by $x=W_1+v_1 t$
(or equivalently $x=W_2+v_2 t$).
\end{itemize}
This procedure gives, for each $r_1\in [-1, \,r_{\sss\rm R}]$ and $t$,
the value of $r_2$ and $x$. In practice, it makes it possible to
associate to each $(x,t)$ a couple $(r_1,r_2)$. The result is shown in
Figs.~\ref{fig3} for the two initial profiles \eqref{ek1} and
\eqref{ek37}.

\begin{figure*}[hptb]
\includegraphics*[width=1.9\columnwidth]{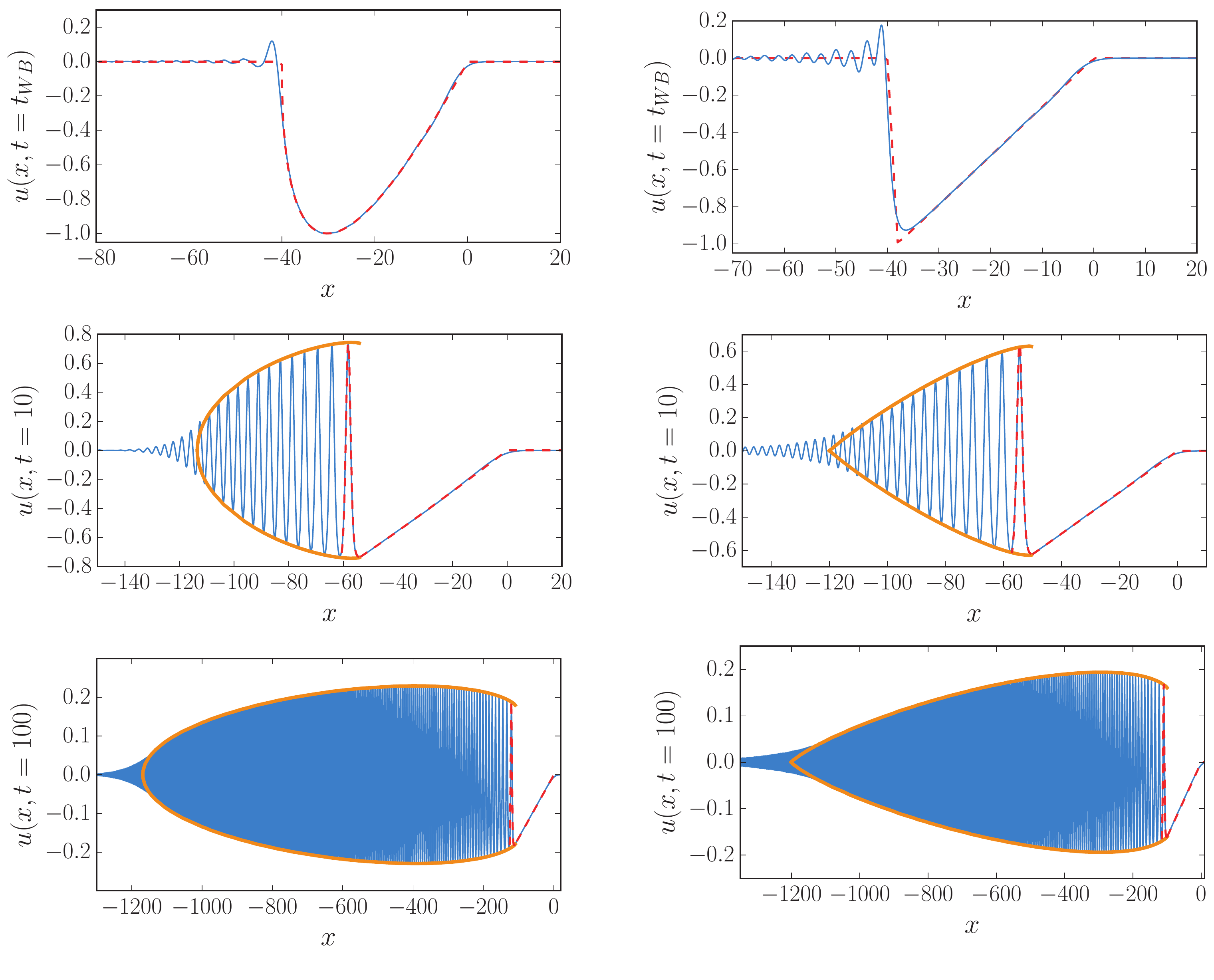}
\caption{$u(x,t)$ as a function of $x$ for fixed $t$. The upper row
  corresponds to the wave-breaking time $t_{\sss\rm WB}$, the central
  row to $t=10$ and the lower one to $t=100$. The left column refers
  to the parabolic initial profile, and the right one to the
  triangular initial profile. The blue solid line corresponds to the
  numerical solution of Eq.~\eqref{eq2}. The envelopes correspond to
  the results of Whitham modulation theory. The dashed red lines
  represent the dispersionless profile $r(x,t)$ and also (in the two
  lower rows) the Whitham result for the soliton at the large
  amplitude boundary of the DSW.}\label{fig4}
\end{figure*}

Note that the characteristics of the DSW are different for the initial
profiles \eqref{ek1} and \eqref{ek37}: for the parabolic profile, in
the upper left plot of Fig. \ref{fig3}, the edge point of the 
DSW---at $(x_0,t_{\sss\rm WB})$---pertains to region A and corresponds
to $r_1=0$, while for the triangular profile, in the lower left plot
of Fig.~\ref{fig3}, the edge point of the DSW belongs to region B,
with $r_1=-1$. For the parabolic profile, the value $r_1=-1$ defines a
line which separates the regions A and B of the plane $(x,t)$ (see the
upper left plot of Fig.~\ref{fig3}). This line reaches a boundary of
the DSW only at $x_{\sss\rm R}(t_{\sss\rm A/B})$, where $t_{\sss\rm
  A/B}$ is the time where part A of the initial profile has just been
completely absorbed within the DSW. On the other hand, for the
triangular profile, the whole left boundary of the DSW corresponds to
the line $r_1=-1$ (see the lower left plot of Fig.~\ref{fig3}).

\

The knowledge of $r_1(x,t)$ and $r_2(x,t)$ makes it possible to
determine, for each time $t> t_{\rm\sss WB}$, $u(x,t)$ as given by the
Whitham approach, for all $x\in\mathbb{R}$:
\begin{itemize}
\item[(i)] In the regions $x\ge 0$ and $x\le x_{\sss\rm L}(t)$, we
  have $u(x,t)=0$.

\item[(ii)] In the region $[x_{\sss\rm R}(t),0]$ $u(x,t)=r(x,t)$ which is
solution of the Hopf equation (obtained by the method of characteristics).

\item[(iii)] Inside the DSW, for $x\in[x_{\sss\rm L}(t),x_{\sss\rm R}(t)]$,
the function $u(x,t)$ is given by the expression
\eqref{ek9}, with $r_3=0$ and $r_1$ and $r_2$ determined as
functions of $x$ and $t$ by the procedure just explained.
\end{itemize}

The corresponding profiles are shown in Fig.~\ref{fig4} for the
parabolic and triangular initial distributions. The agreement with the
numerical simulation is excellent in both cases.

\begin{figure}
\includegraphics*[width=\linewidth]{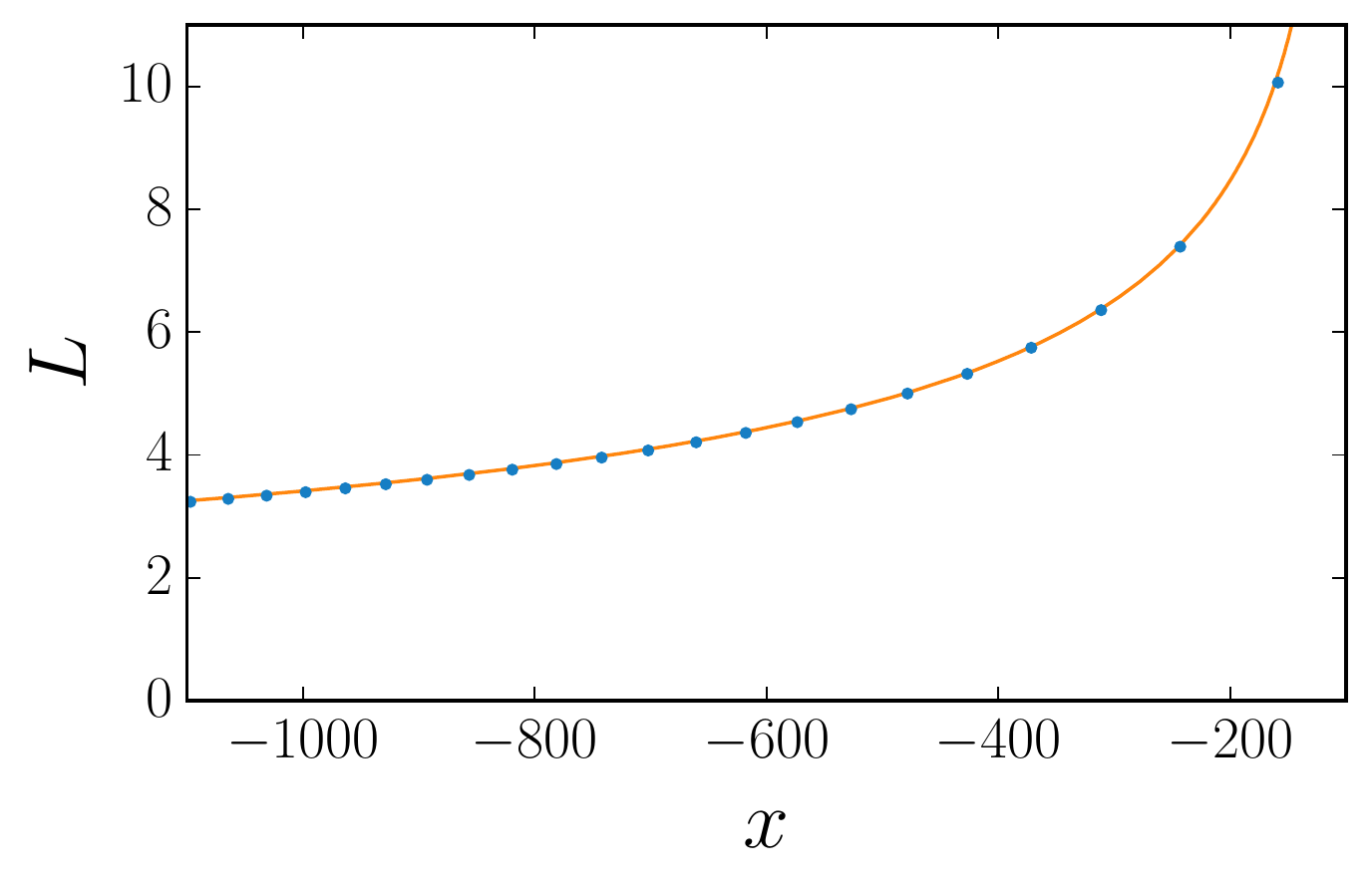}
\caption{Evolution of the wavelength of the nonlinear oscillations within the
  DSW as a function of position $x$. The figure corresponds to the
  time evolution of the parabolic initial profile represented in the
  lower left plot of Fig.~\ref{fig4} ($t=100$). The continuous line
  represents the results of Whitham theory and the points are the
  value of the wavelength extracted from the numerical
  simulations.}
  \label{fig:WL}
\end{figure}

In Fig.~\ref{fig:WL} we also compare the wave-length of the
nonlinear oscillations within the DSW as determined by Whitham
approach [Eq.~\eqref{ek16}]
with the results of numerical simulations, and the agreement
is again very good.

\subsection{The initial square profile}\label{sec:square}

In this section we discuss another type of initial condition, which
we denote as ``square profile'':
\begin{equation}\label{sq1}
u_0(x)=\begin{cases}
-1 & \mbox{for}\quad -x_0\le x \le 0\; , \\
0 & \mbox{elsewhere.}
\end{cases}
\end{equation}
El and Grimshaw already theoretically studied the same initial
condition by using the method just exposed \cite{EG2002}. We will here
compare the theory with numerical simulations to indicate some limitations
of the one-phase Whitham method which we use in the present work.

For this initial profile, wave
breaking occurs instantaneously, and until $t\le x_0/4$
a plateau (i.e., a segment with constant
$u(x,t)=-1$) separates the DSW (at the right) from
a rarefaction wave (at the left). In this configuration, the
DSW corresponds to the standard Gurevich-Pitaevskii scheme
for a step-like initial profile with a
single varying Riemann invariant ($r_2$ in this case). This
DSW can be described using the self-similar variable
$\zeta=(x+x_0)/t$, in this case Eq.~\eqref{ek14} for $i=2$ reads
$\zeta=v_2(-1,r_2)$. One also obtains $x_{\sss\rm L}(t)=-x_0-2\, t$,
$x_{\sss\rm R}(t)=-6\, t$ and the rarefaction wave corresponds to
$r(x,t)=x/6t$ for $x\in[x_{\sss\rm
  R}(t),0]$.

It is interesting to remark that the Gurevich-Pitaevskii DSW can also
be described within the approach exposed in Secs.~\ref{sec:GHM} and
 \ref{sec:global}, by solving Eq.~\eqref{ek17} for $i=2$. Here
$W_2^{\sss\rm A}$ should be computed from
\begin{equation}\label{sq2}
\mathscr{W}^{\sss\rm A}(r_1,r_2)=-x_0
\end{equation}
by means of Eq.~\eqref{ek19}. The form \eqref{sq2} of
$\mathscr{W}^{\sss\rm A}$ comes from \eqref{ek28} with
$w^{\sss\rm A}(r)=-x_0$.

At $t=x_0/4$ the plateau disappears, and one enters in region B with
two varying Riemann invariants. Formulae \eqref{ek29} and \eqref{ek31}
lead here to
\begin{equation}
\begin{split}
\mathscr{W}^{\sss\rm B}(r_1,r_2)=& -x_0-\frac{x_0}{\pi}
\int_{-1}^{r_1} \!\frac{\sqrt{\mu+1} \; d\mu}{\sqrt{-\mu(r_1-\mu)(r_2-\mu)}},\\
=& -x_0+\frac{2\,x_0/\pi}{\sqrt{-r_1(1+r_2)}}\, \times \\
& \times\left\{\Pi\left(\frac{1+r_1}{r_1},m\right)-K(m)\right\},
\end{split}
\end{equation}
where $m=(1+1/r_1)/(1+1/r_2)$ and $\Pi$ is the complete
elliptic integral of the third kind.

\begin{figure}[h!]
\includegraphics*[width=\linewidth]{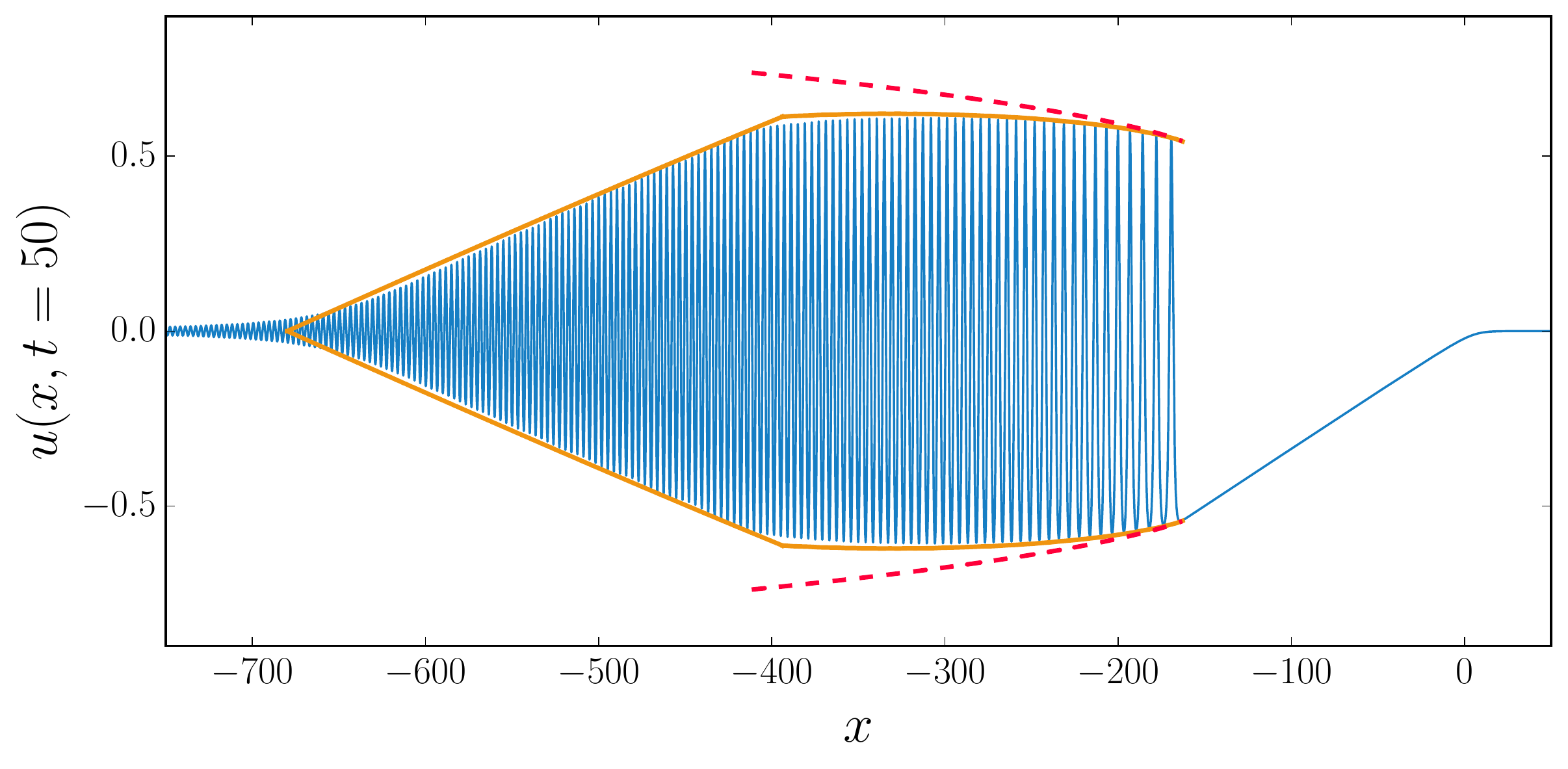}
\includegraphics*[width=\linewidth]{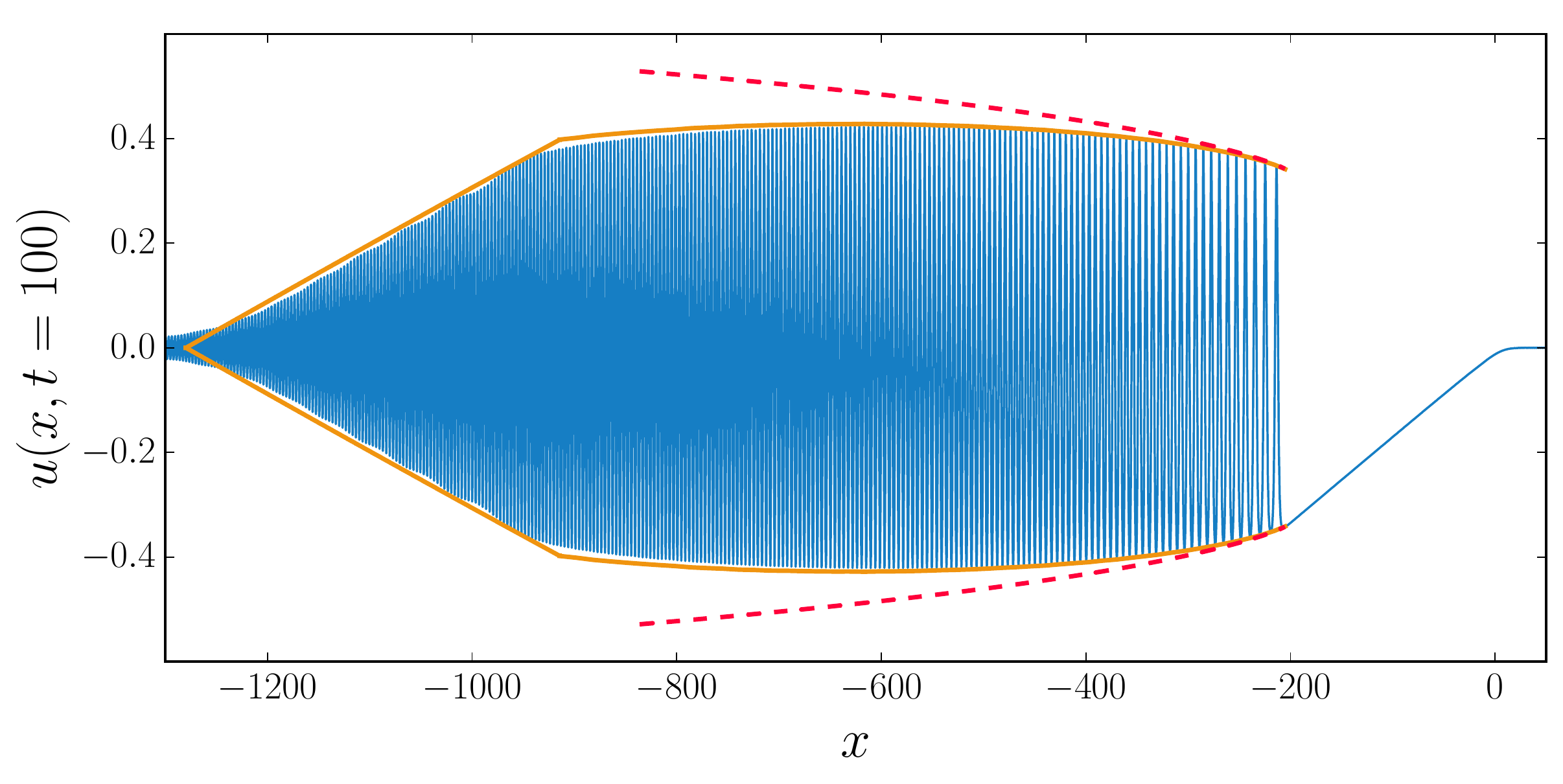}
\caption{Evolution of an initial square profile of type \eqref{sq1}
  with $x_0=80$ after a time $t=50$ (upper plot) and $t=100$ (lower
  plot).  The blue solid lines are the results of numerical
  simulations. The orange envelopes are determined by Whitham method. The
  red dashed envelopes are the asymptotic self-similar results obtained
  in Sec.~\ref{selfsim}. Note the change of scale in the axis of the two
  plots.}
  \label{fig:square}
\end{figure}

The predictions of Whitham theory are compared in Fig.~\ref{fig:square}
with numerical simulations.
Surprisingly enough, the agreement between simulation and theory
decreases at large time: at $t=100$ one can notice oscillations in the
envelope of the front part of the DSW (Gurevich-Pitaevskii
part). Inspection of the dynamics of formation of the nonlinear
structure reveals that, during the formation of the rear rarefaction
wave, some oscillations appear due to dispersive effects associated
with the discontinuity of the initial condition \eqref{sq1}: their
interference with the oscillations of the DSW leads to the modulated
structure which can be observed in the lower plot of
Fig.~\ref{fig:square}. Such a behavior requires a two-phase approach
for a correct description. Note also that for numerical purposes the
initial condition is smoothed \footnote{In the numerical simulations
presented in Fig.~\ref{fig:square} we take
$u_0(x)=\tfrac14 (\tanh(x/\Delta)-1)\times(1+\tanh((x+x_0)/\Delta))$
with $\Delta=2$. This profile tends to \eqref{sq1} when
$\Delta\to 0$.}  and that the beating phenomenon increases for
sharper initial condition (or a lower values of $x_0$).

The predictions of the self-similar solution of Sec.~\ref{selfsim} are
also displayed in Fig.~\ref{fig:square}. In this figure, the envelopes
of the DSW expected from Eqs.~\eqref{39.2} and \eqref{t3-150.20} are
represented by red dashed lines.  In the vicinity of its soliton edge,
the DSW is accurately described by the similarity solution. However
this approach is not able to tackle the other, small amplitude,
boundary of the shock. This is expected since ---as discussed above---
in the small amplitude region a scaling different from the one of Eq.
\eqref{eq4} holds, with the relevant self-similar parameter
$\zeta=(x+x_0)/t$; see Refs.~\cite{dvz-94,Seg81} for a general
discussion.

\section{Conclusion}\label{conclusion}

In the present work we have studied asymptotic solutions of the KdV
equation for which no soliton is formed in the limit $t\to\infty$.
We used the Whitham
modulation theory combined with the generalized hodograph method for
describing the DSW which is formed after wave breaking.
A simple similarity description has been also obtained near the
large amplitude region of the shock, still within the framework of
Whitham's approach. Our results confirm, simplify and extend in some
respects the previous works on this subject. We show that this theory
provides a practical tool for the description of nonlinear evolution of
pulses and can be used for comparison with experimental data. Besides
that, it yields simple enough analytic formulae for some
characteristic features of DSWs and reveals different scaling regimes
of DSW evolution. Extensions of this approach to non completely
integrable equations \cite{Kam18} and to other systems of physical
interest are under study.

\begin{acknowledgments}
  A. M. K. thanks Laboratoire de Physique Th\'eorique et Mod\`eles
  Statistiques (Universit\'e Paris-Saclay) where this work was
  started, for kind hospitality. This work was supported by the French
  ANR under Grant No. ANR-15-CE30-0017 (Haralab project).
\end{acknowledgments}

\end{document}